\documentclass[11pt]{article}

\renewcommand{\baselinestretch}{1.2}
  \renewcommand{\arraystretch}{1.0}
 \usepackage{amsmath}
\usepackage{amscd}
\usepackage{amssymb}
\usepackage{graphicx}
\usepackage{ifpdf}
\usepackage{fancyhdr}

\begin{document}

 \title{A Note on the Bellare-Rivest Protocol for\\  Translucent Cryptography }
  \author{Zhengjun Cao$^{1}$, \qquad Lihua Liu$^{2,*}$} 
  \footnotetext{ $^1$Department of Mathematics, Shanghai University, Shanghai,
  China. \quad      $^2$Department of Mathematics, Shanghai Maritime University,   Shanghai,
  China.   $^*$\,\textsf{liulh@shmtu.edu.cn}    }

\date{}
\maketitle

\begin{abstract}
We remark that the Bellare-Rivest protocol for translucent cryptography [J. Cryptology (1999) 12: 117-139] can not truly enable
 the government to decrypt partial encrypted communications.

 \end{abstract}

\section{Introduction}
The primitive of translucent cryptography introduced by Bellare and Rivest \cite{BR99} is viewed as an alternative
to the controversial key-escrow techniques. It aims to achieve an appropriate balance between individual privacy and government
access to communications. The main idea behind their protocol for the primitive can be described as follows.

 Suppose that Bob's public key is $pk_{B}$ and Larry's public parameters are
$V_1, \cdots, V_t $, where the role of Larry is played by the government. Among these $V_1, \cdots, V_t $, only one is the good parameter for Larry, i.e., he knows the relevant  trapdoor. But nobody knows which is the good parameter except Larry himself.
When Alice wants to send a session key $s$ to Bob, which shall  be used for the later symmetric encryption, she randomly picks a parameter $V_i\in \{V_1, \cdots, V_t\}$ and a secret exponent $k$  and sends  $\{g^k, sy_B^k;  s{V_i}^k, i \}$, where $g$ is the generator of the ElGamal cryptosystem and $y_B$ is Bob's public key. If $V_i$ is just the good parameter, then Larry shall successfully recover $s$. Otherwise, Larry fails to recover it.

The Bellare-Rivest protocol is
based on noninteractive fractional oblivious transfer  and uses just one ElGamal encryption. In the scenario, Larry plays the role of receiver. But we find the OT model is not appropriate to this situation because: 1)
   \emph{the transferred messages is not recognizable for the receiver}, 2) \emph{the sender is not willing to disclose some messages to the receiver}. Moreover, there is short of a mechanism for ElGamal encryption to force the sender to invoke any $V_i\in \{V_1, \cdots, V_t \}$. That means the sender can chose an index $i$, a secret exponent $k$ and a
random element $\hat{V}$ and generate the malformed ciphertext $\{g^k, sy_B^k;   s{\hat V}^k, i \}$. In such case, Bob can successfully recover the session key $s$,  but Larry can not recover it. Thus the Bellare-Rivest protocol for translucent cryptography can not truly enable
  government to decrypt partial encrypted communications.

  \section{Description of the Bellare-Rivest protocol}

  The implementation of translucent cryptography based on noninteractive
oblivious transfer will not incur any ``extra flows". When Alice wishes to communicate
with Bob, her only transmission is to Bob.  If the government wants to know something about what
Alice is saying to Bob, it must wiretap their communications, and then it will be able to
decrypt a fraction  of the messages it picks up.

\emph{Global setup}. The ACLU (American Civil Liberties Union) picks a large
global prime $\rho$ (say at least 1024 bits in length),
a generator $g$ of the multiplicative group $\mathbb{Z}_{\rho}^*$  and a value $U$ such that no one
knows the discrete logarithm $\log_g (U)$  of $U$, modulo $\rho$.

\emph{Publication of Larry's public parameters}. Larry picks a secret exponent $x_L\in \mathbb{Z}_{\rho-1}$ and a random index $\ell\in\{1, \cdots, t\}$,  and computes $V_{\ell}=g^{x_L} \mod \rho,$
$V_j= V_{\ell}U^{j-\ell} \mod \rho, j=1, \cdots, t$.  Notice that $V_{\ell}$ is the only one good key for Larry.   Only Larry knows its secret index $\ell$.
Any one can verify that whether $V_1, \cdots, V_t$ are well-formed by checking $V_j/V_1=U^{j-1}\mod \rho$, where $j \in \{2, \cdots ,t\}$.

\emph{Encryption.} Suppose that Bob's public key is $y_B=g^{x_B}\mod \rho$, for some secret exponent $x_B\in \mathbb{Z}_{\rho-1}$.  When Alice wants to send a session key $s$ to Bob, she picks a random exponent $k \in \mathbb{Z}_{\rho-1}$ and $V_i\in \{ V_1, \cdots, V_t\}$, and sends
the ciphertext $\{c_1, c_2;  c_3, i\}=\{g^k, sy_B^k; sV_i^k, i \}$.

\emph{Decryption for Bob.} Bob computes $s=c_2/c_1^{x_B} \mod \rho$.

\emph{Decryption for Larry.} If $V_i$ is the good key for Larry, he computes $s=c_3/c_1^{x_L} \mod \rho$.

\section{Remarks on the Bellare-Rivest protocol}

From the practical point of view, the Bellare-Rivest Protocol has some drawbacks.

(1) \emph{Reveal of good key}.  It is easy to find that  Larry's security relies on the fact that Alice does not know his secret index $\ell$; if she did, she could encrypt using only the other keys, and Larry would never be able to recover the session key $s$. Suppose that Larry needs to use the wiretap
information as evidence in a court case. The plaintexts are then revealed, and, by
their examination, Alice can determine which of her messages were decrypted. This tells
her what Larry's secret index is.

Bellare and Rivest \cite{BR99} suggest that Larry have many public keys, with different keys used in different programs or devices at different times.
Clearly, this will incur heavy cost for key management. This drawback renders the protocol unrealistic.

(2) \emph{Malformed ciphertext attack}. In order to enable Larry (the government) to decrypt partial encrypted communications, Alice must follow the translucent cryptography
protocol properly. If Alice generates the ciphertext as $\{g^k, sy_B^k; s{\hat V}^k, i \}$ for some random element $\hat V\notin \{V_1, \cdots, V_t\}$ and $i\neq \ell$, then
Larry can not find her trick. Even if $i= \ell $, Larry can not recover the session key $s$.

 Bellare and Rivest \cite{BR99} suggest that Larry's key should be digitally signed under some global public key which is embedded in a commercially available crypto-box. Note that
Larry's key cannot be embedded in the box because they may need to be changed now and then. Regretfully, they did not specify the method to force Alice to invoke Larry's key.
In fact, it is impossible for  ElGamal cryptosystem to check the  ciphertext $\{g^k, sy_B^k; sV_i^k, i \}$ is well formed, i.e., $V_i$ is indeed invoked and included in the third component. In short, the attack discounts greatly the importance of the protocol.

\section{Misused noninteractive fractional oblivious transfer}

 The oblivious transfer primitive is due to Rabin \cite{R81}.
It has been formalized and extended by  Even et al. \cite{EGL85}.
We pint out that in most reasonable applications of OT,
   \emph{the transferred messages must be recognizable for the receiver},
    or \emph{the sender is willing to disclose some messages to the receiver}.  The property has been explicitly specified in the earlier works. We refer to the following descriptions.

 In Ref.\cite{R81}, Rabin explained  that:

 \begin{quotation}
  Bob and Alice each have a secret, SB and SA, respectively, which they wish to exchange. For
example, SB may be the password to a file that Alice wants to access (we shall refer to this file as
Alice's file), and SA the password to Bob's file. To exclude the possibility of randomizing on the possible digits of the password, we assume
that if an incorrect password is used then the file is erased, and that Bob and Alice want to guarantee
that this will not happen to their respective files. \end{quotation}

 In Ref.\cite{EGL85}, Even, Goldreich and Lempel  stressed  that:  \begin{quotation}

The notion of a ``recognizable secret message" plays
an important role in our definition of OT. A message is
said to be a recognizable secret if, although the receiver
cannot compute it, he can authenticate it once he receives
it.

The notion of a recognizable secret message is evidently
relevant to the study of cryptographic protocols,
in which the sender is reluctant to send the message
while the receiver wishes to get it. In such protocols, it
makes no sense to consider the transfer of messages that are
either not secret (to the receiver) or not recognizable (by the
receiver). \end{quotation}

In symmetric case, such as exchanging secrets, signing contracts,  both two participators can easily verify the correctness of the received messages.
 In unsymmetric case, such as a database manager plays the role of the sender and a client plays the role of the receiver,
it is usual that the sender is willing to disclose some messages to the receiver.

It claims that the Bellare-Rivest protocol is based on noninteractive oblivious transfer, which is due to Bellare and Micali \cite{BM89}.
We here stress that the transferred session key $s$ in their protocol is not recognizable (by the
government). Besides, the sender (Alice) is reluctant to reveal $s$ to  the government (Larry). Thus, the primitive of OT is not appropriate to this case.
Frankly speaking, the essential technique used in the Bellare-Rivest protocol is to mix some padding parameters with the Larry's true parameter (he knows the relevant trapdoor)
such that others can not decide which parameter is his true public key.

\section{Conclusion}

The Bellare-Rivest protocol for translucent cryptography seems impossible to implement although interesting.
 We hope this note is helpful to clarify some misunderstandings about noninteractive oblivious transfer as well as OT itself.

\end{document}